# Giant Rotational Meta-Doppler from Genetically Designed Superscatterers

**Dmytro Vovchuk, Sergey Geyman, Konstantin Grotov, Dmitry Dobrykh, Andrey Machnev, Anna Mikhailovskaya, Susanna Rozental, Anton Kharchevskii, Mikhail Tsukerman, Vjaceslavs Bobrovs, Aviel Glam, Alexander Gumennik, and Pavel Ginzburg**

## ABSTRACT

The motion of a rigid body interacting with a wave leaves spectral signatures, with the Doppler shift as the dominant contribution. Since any motion can be decomposed into translational and rotational components, rotational Doppler provides additional information about the object's dynamics. In the electromagnetic domain, rotating objects generally produce rotational Doppler, or micro-Doppler, signals determined by the symmetry and spatial structure of the scattering process. For objects that are subwavelength or comparable in size to the wavelength, the response is typically dominated by the lowest dipolar scattering channel, so the leading spectral component commonly appears at twice the angular frequency. Here, we introduce the concept of artificially engineered rotational micro-Doppler by designing a compact, strongly scattering structure that operates through a high-order multipolar cascade of resonances, thereby producing a giant enhancement. Magneto-electric arrays composed of strongly coupled electric and magnetic resonators are optimized in the GHz range using a covariance matrix adaptation genetic algorithm to maximize the micro-Doppler frequency. Unlike conventional higher-order multipole designs used in superscatterers for a specific angle of incidence and polarization, our approach jointly optimizes excitation and scattering under radar-relevant conditions for a rotating blade. The resulting arrays exhibit a giant rotational meta-micro-Doppler response, exceeding the dipolar limit by two orders of magnitude and mapping rotations of tens of hertz into the kilohertz range. Beyond its fundamental significance, this mapping has practical value because it shifts rotor micro-Doppler signatures well above slow-moving radar clutter, thereby improving the detectability of slow motion.

**corresponding author (*Dmytro.Vovchuk@rtu.lv*)*

## INTRODUCTION

The Doppler effect is a fundamental wave phenomenon that explains frequency shifts caused by the relative motion between a wave source and an observer. In remote sensing, including radar, the Doppler effect plays a crucial role in determining the velocity and motion of objects by analyzing frequency shifts in the scattered signal. Micro-Doppler effects in radar and optical sensing arise from small-scale motion components such as vibrations, oscillations, and rotations, providing valuable insights into target dynamics [1]. Rotational Doppler, as a special case of micro-Doppler, has been extensively studied across various domains, tracing back to experiments with macroscopic objects in the microwave [2–5] and optical domains[6], as well as spinning molecules[7,8], to name a few.

The frequency shifts induced by rotational motion typically have a dominant contribution at twice the rotor's angular velocity ($2\Omega$), provided the rotor's size is smaller or comparable to the wavelength[3]. In the classical case, this effect is determined by the scatterer's symmetry properties. For example, the electromagnetic interaction of a plane wave with a rotating wire follows the periodicity dictated by the rotor's symmetry group[3]. In radar interrogation, the number of blades directly determines the frequency-multiplication factor in the Doppler spectrum. For example, a three-bladed rotor, with its three-fold rotational symmetry ($C_3$), generates micro-Doppler shifts at $3\Omega$, while a four-bladed rotor, with four-fold symmetry ($C_4$), produces shifts at $4\Omega$. Thus, the number of blades directly determines the frequency multiplication factor in the Doppler spectrum. Similar considerations apply to rotating molecules, where dipolar transitions often dominate. Further enhancements were achieved by scattering complex beams from rotating macroscopic objects larger than the wavelength in the optical domain[9,10].

To achieve a maximal $N\star\Omega$ shift (N-fold enhancement), an appealing approach is to design a scatterer that supports a higher-order resonant multipole (M-pole), thereby exploiting the strong angular dependence of its far-field signature[11]. While this concept is not without merit, it overlooks the excitation problem. The scatterer must remain efficiently excitable at any relative angle to the incident plane wave, which requires sufficient overlap with an excitation field of different symmetry. Moreover, practical considerations arise in designing structures that support high-order resonant multipoles, as they inherently exhibit strong near-field accumulation within their interior, making them more susceptible to losses and fabrication tolerances. Similar constraints arise in related phenomena such as superscattering and superdirectivity, where a large number of multipoles must interfere constructively to achieve either a high scattering cross-section or a highly directive antenna pattern[12].

The design concept for scattering management with multi-resonant structures can be inspired by principles from the field of superscatterers[13–21]. Superscatterers are defined as subwavelength structures that exhibit scattering cross-sections exceeding the maximum theoretical limit of a ***single*** resonant lossless dipole, which is $3\lambda^2/(2\pi)$, where $\lambda$ is the excitation wavelength. This channel limit scales with the multipole order as $\ell(\ell+1)$, where for a dipole, the prefactor is 3 ($\ell=1$). To bypass the single-channel limit, a small scatterer must support several resonant multipoles, which introduces performance-related challenges, including strong near-field accumulation and high sensitivity to losses, ultimately constraining efficiency. Considering these factors, selecting an appropriate material platform for implementation and performing intensive optimization across a large parameter space are essential. Genetically designed electric, magnetic, and magneto-electric arrays combining resonant dipoles - straight metallic wires and split-ring resonators (SRRs) have been proposed as promising candidates[22–25].

Here, we develop flat magneto-electric arrays with a strong differential angle-dependent backscattering cross-section and extreme angular dependence. These one-dimensional and two-dimensional structures, composed of strongly coupled resonators, are designed to be positioned on a rotating plane. When interrogated by radar, they exhibit a maximized frequency shift that significantly exceeds twice the angular rotational frequency, $2\Omega$, which we define as the single-channel (or dipolar) micro-Doppler limit. Maximizing this effect is set as the objective function in the genetic optimization process.

A prospective practical use case illustrating the concept is presented in Figure 1, where a giant-meta-Doppler structure enhances the radar visibility of small drones in urban clutter, thereby providing an additional layer of safety. Although this example establishes an immediate connection to a realistic scenario, the main focus here is on revealing the fundamental capability of generating meta-micro-Doppler responses.

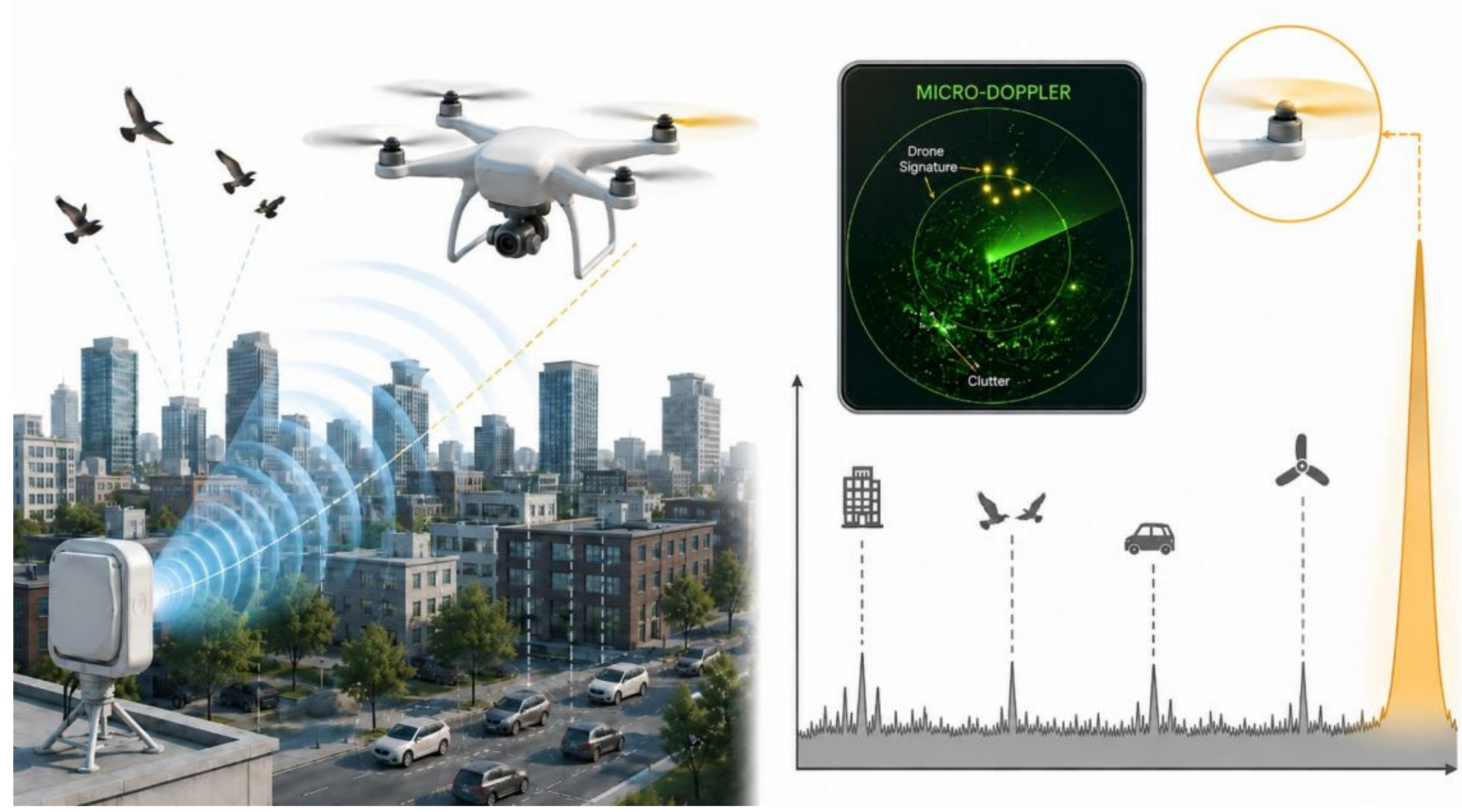

**Figure 1. Practical illustration of a giant meta micro-Doppler in clutter.** A rotating blade carrying an engineered meta-structure generates a bright, high Doppler-shifted radar signature that rises above urban clutter, whereas birds, cars, and other background scatterers contribute mainly at low baseband frequencies.

The study is organized as follows: first, the optimization framework, objective function, and numerical methodology are introduced to define the design problem and the path toward enhanced rotational scattering. Next, the optimized 1D and 2D arrays are analyzed using angular RCS, near-field distributions, and multipolar decomposition to reveal the physical origin of the effect and identify the underlying resonance-cascading mechanism. Having established this basis, direct rotating measurements are then presented to verify that the engineered angular response is indeed converted into enhanced micro-Doppler signatures. Finally, the dependence on frequency and elevation angle is examined to assess the robustness of the phenomenon and its practical applicability.

## METHODS

### *The optimization search space and the cost function*

As a basic design for a rotating scatterer, we consider a flat magneto-electric array, consisting of wires that primarily interact with the electric field component and split-ring resonators (SRRs) that respond to the magnetic field. In both cases, field polarization plays a significant role. For efficient scattering, dipoles are oriented with the electric field, while the axis normal to the rings aligns with the magnetic field component (Figure 2(a-b)). Rectangular SRRs are chosen primarily for the convenience of fast computations, as will be elaborated further. Two main layouts will be considered. The first configuration is a 5x1 one-dimensional (1D) array. This structure includes 5 unit cells, each of which can be populated with either a wire or an SRR. Each element can be rotated by a different angle relative to the array's major axis and scaled, as shown in Figure 2(a). The geometric center of either the dipole or SRR coincides with the center of the unit cell, thereby reducing the potential number of degrees of freedom in the optimization. The second structure to be considered is a 5x5 planar array populated using the same strategy. In this case, the number of independent degrees of freedom is 25 discrete (either wire or ring) and 2x25=50 continuous, related to the scaling and rotation of the structures. Considering the intensive optimization, which requires multiple calculations of the electromagnetic scattering problem (the forward solver), the search space rapidly becomes quite large. In terms of use cases, the 5x1 array could be implemented on a drone's rotating blade, while the 5x5 array offers much greater optimization capabilities, making it more promising for achieving a large micro-Doppler effect when a larger tagging area is available.

The interaction scenario is as follows: a linearly polarized wave is incident on the array, with the normal aligned with the magnetic field and the electric field parallel to the array. The array rotates around its normal axis without precession (Figure 2(b)). The cost (or objective) function is formulated as follows. Without loss of generality, the operating frequency $f$=10GHz is chosen to align with one of the radar bands (X-band) commonly used for surveillance applications. The monostatic radar scattering cross-section (RCS), or backscattering cross-section ($\sigma$), is evaluated at several rotational angles (N = 50), evenly distributed within the $2\pi$-rotation interval, thereby setting the highest uniquely resolvable harmonic to $m_{max}$=25. At the next stage, the angular response is Fourier-transformed, and the optimization objective is set to maximize the amplitudes of the highest accessible harmonics, i.e. $S_m = \frac{1}{N}\sum_{n=0}^{N-1} \sigma(\theta_n)\, e^{-im\theta_n}, \theta_n = \frac{2\pi n}{N}$. This choice is intended to demonstrate the largest achievable micro-Doppler effect, rather than to directly reflect a practical radar scenario, in which robustness to the relative orientation between the rotor plane and the radar, together with reproducibility of the rotational signature, and other constraints may be more important than maximizing the Doppler enhancement itself. It is worth noting that maximizing the highest detectable micro-Doppler component, regardless of its amplitude, is ill-posed, as the optimizer may identify arbitrarily high-order spectral features that are physically negligible. To avoid this issue, the objective function was defined as a composite score that simultaneously rewards large frequency multiplication, strong amplitude, and pronounced angular contrast:

$$J = m \cdot |S_m| \cdot \sigma_{\text{std}} \,, \tag{1}$$

*m* is the index of the highest frequency component among the k=3 largest amplitude peaks in the Fourier spectrum, and $\sigma_{\text{std}}$ denotes the standard deviation of the RCS angular dependence. The parameter k restricts the search to the dominant spectral components, preventing the selection of high-order features caused by numerical noise. While k=1 favors only the strongest and typically low-order response, larger values increase sensitivity to weak, physically irrelevant terms. Thus, k=3 provides a robust balance between amplitude significance and frequency multiplication.

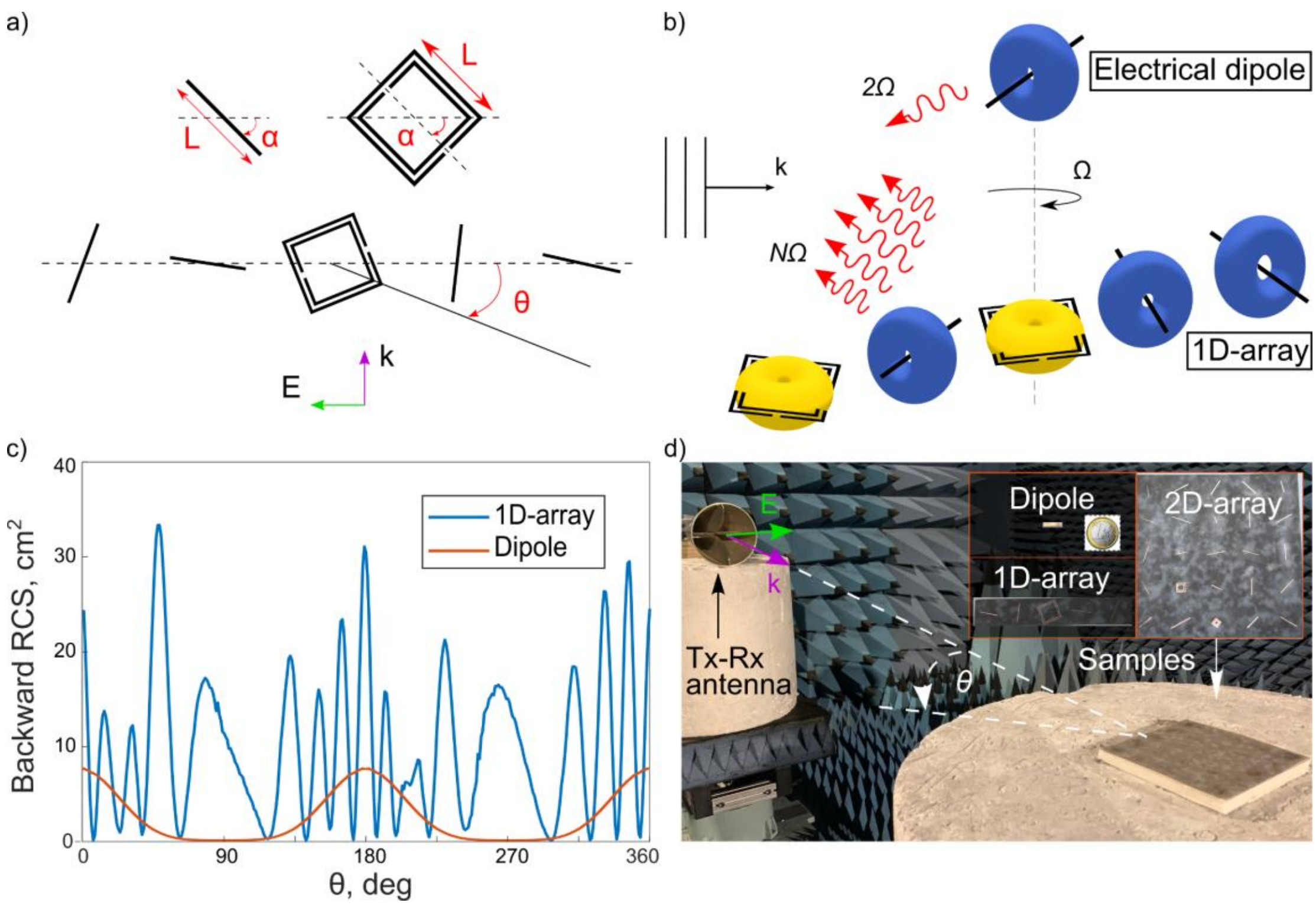

**Figure 2. Concept and experimental configuration of the meta micro-Doppler scatterer.**

**(a)** Basic elements used in the optimization and their variable parameters: length L and angle $\varphi$, together with the geometry of the optimized 1D array. **(b)** Schematic illustration of meta micro-Doppler generation in an array composed of electric dipoles and split ring resonators. Dipolar scattering diagrams are shown in blue for electric resonators and in yellow for magnetic resonators. **(c)** Numerically calculated angular dependence of the RCS for the optimized 1D array and a resonant electric dipole at 10 GHz. **(d)** Experimental setup with a rotating table for RCS measurements of the three samples shown in the inset.

*The optimization algorithm*

Given the large search space and the highly resonant nature of the electromagnetic problem, genetic optimization is the preferred approach. Optimization algorithms can be classified into two main categories based on whether they utilize objective function gradients[26] or not[27–29]. The choice of the most suitable algorithm depends on the specifics of the problem. Based on previously reported considerations[30], a covariance matrix adaptation evolution strategy (CMA-ES)[31] was used. Unlike traditional genetic algorithms (GAs), particle swarm optimization (PSO), differential evolution (DE), or simulated annealing (SA), CMA-ES is especially effective for high-dimensional, multimodal optimization problems and requires minimal tuning. Its adaptive covariance matrix facilitates efficient exploration of the search space and accelerates convergence. With its data-driven mutation and implicit crossover, CMA-ES is particularly suited for complex electromagnetic problems where the objective function landscape is highly irregular[32].

*The forward solver and justification*

Given that hundreds of iterations are needed for a genetic algorithm convergence, an efficient electromagnetic solver is essential. Scattering from shaped thin metal wires can be analyzed using the Hallén or Pocklington integral formulations, which reformulate the problem to significantly reduce its dimensionality[33]. These methods use the Method of Moments (MoM), in which the wires are discretized into segments, each represented by a basis function. The integral equations governing the system are transformed into matrix equations. Solving these provides the current distribution along the wires, allowing the calculation of the resulting scattered fields. This approach provides a computationally efficient framework for analyzing scattering from curved wires, with substantially shorter runtimes than the finite element method (FEM) and the finite-difference time-domain (FDTD) techniques typically implemented in commercial software. The PyNEC Python package[34], based on NEC-2, has been used as a method implementation. To enhance the efficiency of the scattering calculations, we incorporated parallel computation at each iteration of the CMA-ES evolutionary algorithm. In this approach, the scattering calculations for individual realizations in the population are distributed across available CPU processes. This parallelization enables the simultaneous evaluation of scattering for multiple structures, greatly accelerating computation.

After the optimizer ran, full-wave simulations were performed in CST Microwave Studio using the time-domain solver. A linearly polarized plane wave served as the source. All elements were composed of square-section wires, 0.5 mm thick, made of a perfect electric conductor (PEC). The rotation angle was varied in $1°$ increments. A finer angular resolution was used in CST to avoid aliasing, accurately capture narrow scattering features, and ensure reliable extraction of higher-order spectral components (Figure 2(c) for a representative example).

*Fabrication and Experimental Details*

The structures were fabricated with a conventional printed circuit board (PCB) methodology. To minimize the effect of a substrate's permittivity on the resonant response, the sample was fabricated by chemically etching copper elements on a very thin dielectric board: Rogers 3003 ($\varepsilon_r = 3$, $\tan(\delta) = 0.001$) with a thickness of 0.127 mm.

RCS measurements were performed in an anechoic chamber using a broadband horn antenna (NATO IDPH-2018, 2-18 GHz) connected to a Keysight network analyzer (N5232B PNA-L). The horn served as both transmitter and receiver. Quantitative calibration was performed with a 10 cm brass disk, and time gating was applied in post-processing to suppress multipath contributions.

To probe the angular dependence of the operating regime, near-field electric and magnetic maps were measured under the same illumination conditions at 10 GHz (Figure 2(d)). The horn antenna was positioned 1.5 m from the sample and connected to port 1 of a microwave network analyzer. An electric or magnetic probe mounted on a Midas scanning system from ORBIT/FR Engineering Ltd. was raster-scanned 5 mm above the structure and connected to port 2 via a 35 dB low-noise amplifier (LNA, MWA020180-1-4019, 2–18 GHz) and a bandpass filter (VHF-6010+, 6.3–15 GHz). The scanned area was $120 \times 120$ mm² with a 1 mm step for the 1D scatterer and $130 \times 130$ mm² with the same step for the 2D scatterer. After subtraction of the free space reference, the dominant electric field component $E_z$ and the tangential magnetic field $H_t = \sqrt{H_x^2 + H_y^2}$ were compared with numerical simulations.

*Multipolar Analysis*

The multipole analysis was performed numerically using a surface-integration approach[35]. The scattered electric field was exported on a spherical surface enclosing the structure and projected onto vector spherical harmonics using Lebedev quadrature with 5810 points, yielding the electric and magnetic multipole coefficients $a_{lm}$ and $b_{lm}$. The total scattering cross section was then reconstructed from the individual contributions up to $l = 8$. The procedure was validated by comparing the reconstructed total cross section with direct Poynting vector integration.

## RESULTS

### *1D Array*

The first configuration under consideration is a 1D 5x1 array. The rationale for selecting this layout is its potential application in labeling airborne targets[36]. The basic elements of the array are shown in Figure 2(a), whereas the full structure is presented in Figure 2(b), where the scattering diagrams of the electric and magnetic dipoles are superimposed on the corresponding resonators. The collective response of these elements is responsible for the effect. The angular-dependent RCS of the optimized 5x1 array is shown in Figure 2(c), along with a comparison to that of a single rotating dipole. The optimized structure yields a fourfold increase in differential RCS compared to the dipole.

Figure 2(d) shows the experimental setup used for the measurements, whereas the insets present the fabricated structures and indicate the relevant length scales. From this point onward, the numerical predictions and experimental results are considered together. Figure 3(a) shows the angular RCS spectra obtained from CST calculations and measurements, demonstrating a high level of agreement.

Figure 3(b) shows the Fourier-transformed angular spectra of the reference dipole, the 1D array, and the 2D array. Although the 2D structure is discussed later, this comparison already indicates the advantage of using a larger number of elements, both in terms of differential RCS and harmonic order. Inspection of the results already shows that the 25th harmonic of the optimized 1D array carries an amplitude comparable to that of the dipole response.

To reveal the physical mechanism underlying the optimized structure's operation, the surface currents excited on the resonators are analyzed, along with the scattering diagrams shown in Figure 3. Several representative angles, corresponding to the peaks on the angular-dependent RCS plot – 0°, 78°, and 192°, have been selected for the analysis. By comparing performance at these different angles, it can be observed that the peaks in each case correspond to the excitation of different combinations of resonators within the array (Figure 3(c), top row). This observation supports the resonance-cascading principle, in which different resonances in the array are activated under different excitation scenarios. A similar conclusion can be drawn from the scattering diagrams, which exhibit a large number of lobes with strong angular dependencies (Figure 3(c) – the bottom row). An exact multipolar analysis of the scattering diagram will be proved hereinafter.

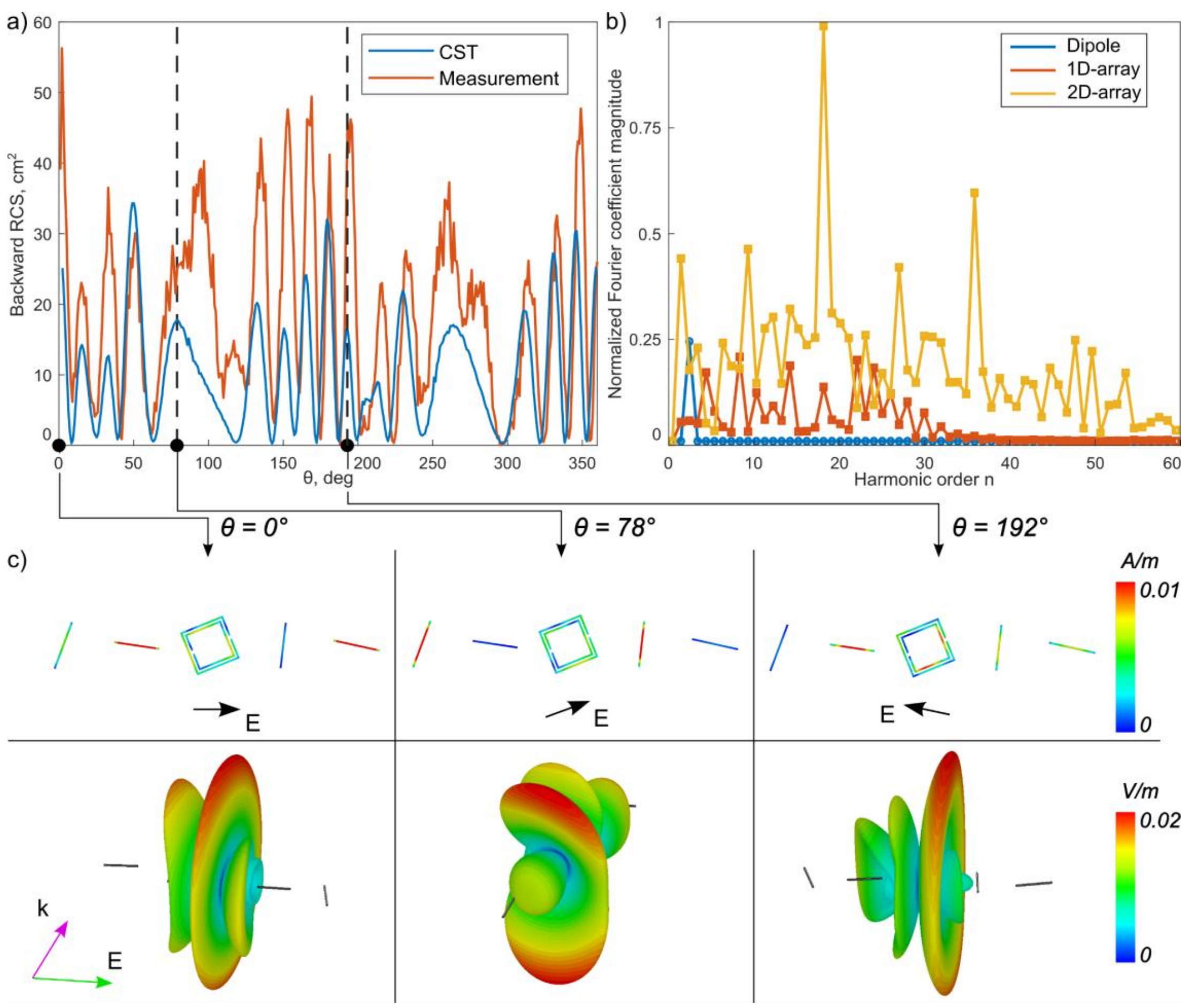

**Figure 3. Numerical and experimental validation of the optimized 1D array. (a)** Angular RCS spectra obtained from CST calculations and measurements at 10 GHz. **(b)** Fourier transformed angular spectra of the reference dipole, optimized 1D array, and optimized 2D array. The 2D result is included for comparison and is discussed in detail below. **(c)** (Top row) Surface currents, and (bottom row) far field electric patterns, calculated at 10 GHz for rotation angles of 0°, 78°, and 192°.

To further verify the physical mechanism inferred from the angular RCS spectra and surface current distributions, the near-field response of the optimized 1D structure was analyzed. Numerical calculations were performed in CST Microwave Studio using E- and H-field monitors. Experimental maps were acquired with the Midas scanning system from ORBIT/FR Engineering Ltd. using electric and magnetic probes at a height of 5 mm above the structure plane, as described in Methods. The results, summarized in Figures 4-5, show that the dominant electric field component is oriented along the z axis, whereas the magnetic response is primarily the tangential component. Consistent with the numerical analysis, the measurements show that different resonators are activated at different incidence angles.

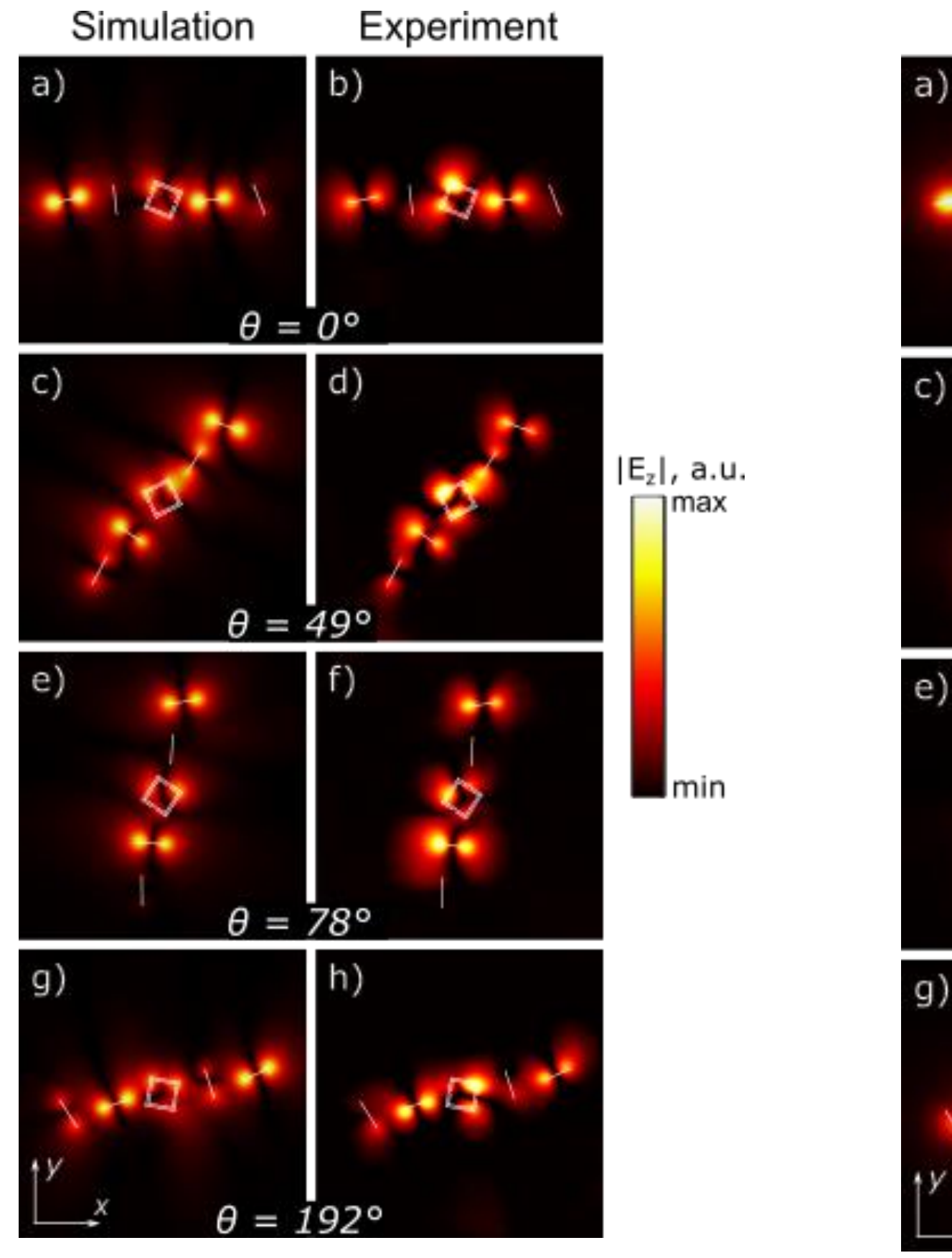

**Figure 4. Near-field electric distributions of the optimized 1D array under rotation.** Numerical results (left column) and experimental maps (right column) are shown for four representative rotation angles: 0°, 49°, 78°, and 192°.

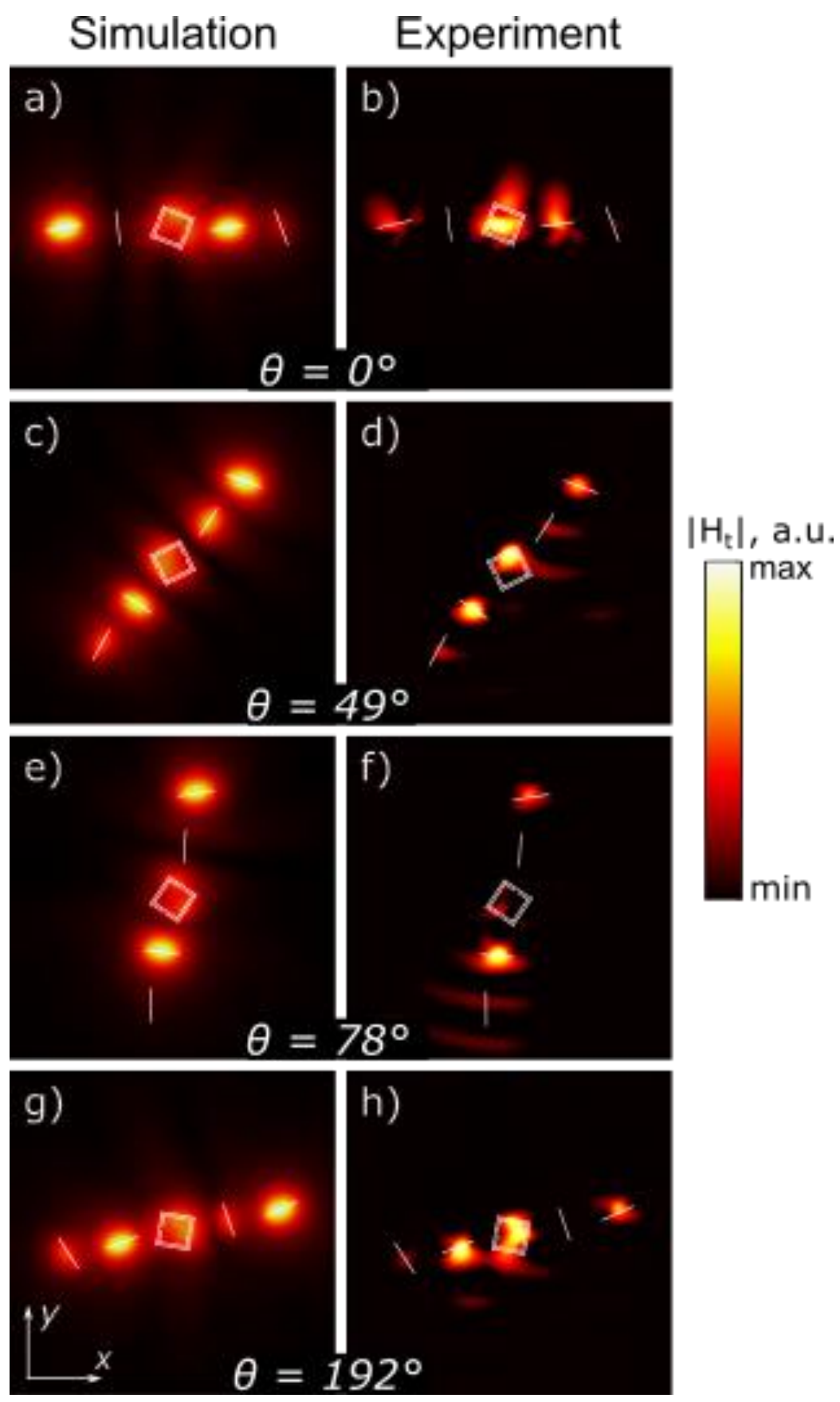

**Figure 5. Near-field magnetic distributions of the optimized 1D array under rotation.** Numerical results (left column) and experimental maps (right column) are shown for four representative rotation angles: 0°, 49°, 78°, and 192°.

To reveal the resonance cascading mechanism, the total scattering cross section of the 1D array was decomposed into spherical multipoles up to the 8th order. The analysis was carried out for four rotation angles corresponding to the cases discussed in Figures 3, 4 and 5. The results, summarized in Figure 6, show the relative contributions of each multipolar term to the total scattering, with the color scale indicating their weights at a given orientation. The column sums are close to 100%, indicating that the dominant multipolar contributions to the scattering are properly accounted for. At each selected angle, the scattering is dominated by a different multipolar contribution, each associated with a strongly angle-dependent radiation pattern.

Higher-order terms contain many lobes, which results in a pronounced variation under rotation. For clarification, in structures lacking rotational symmetry, internal resonances and resonant multipolar contributions are not in 1:1 correspondence. A given excited resonance of the structure may generate a far-field scattering response composed of several multipolar terms[25,37].

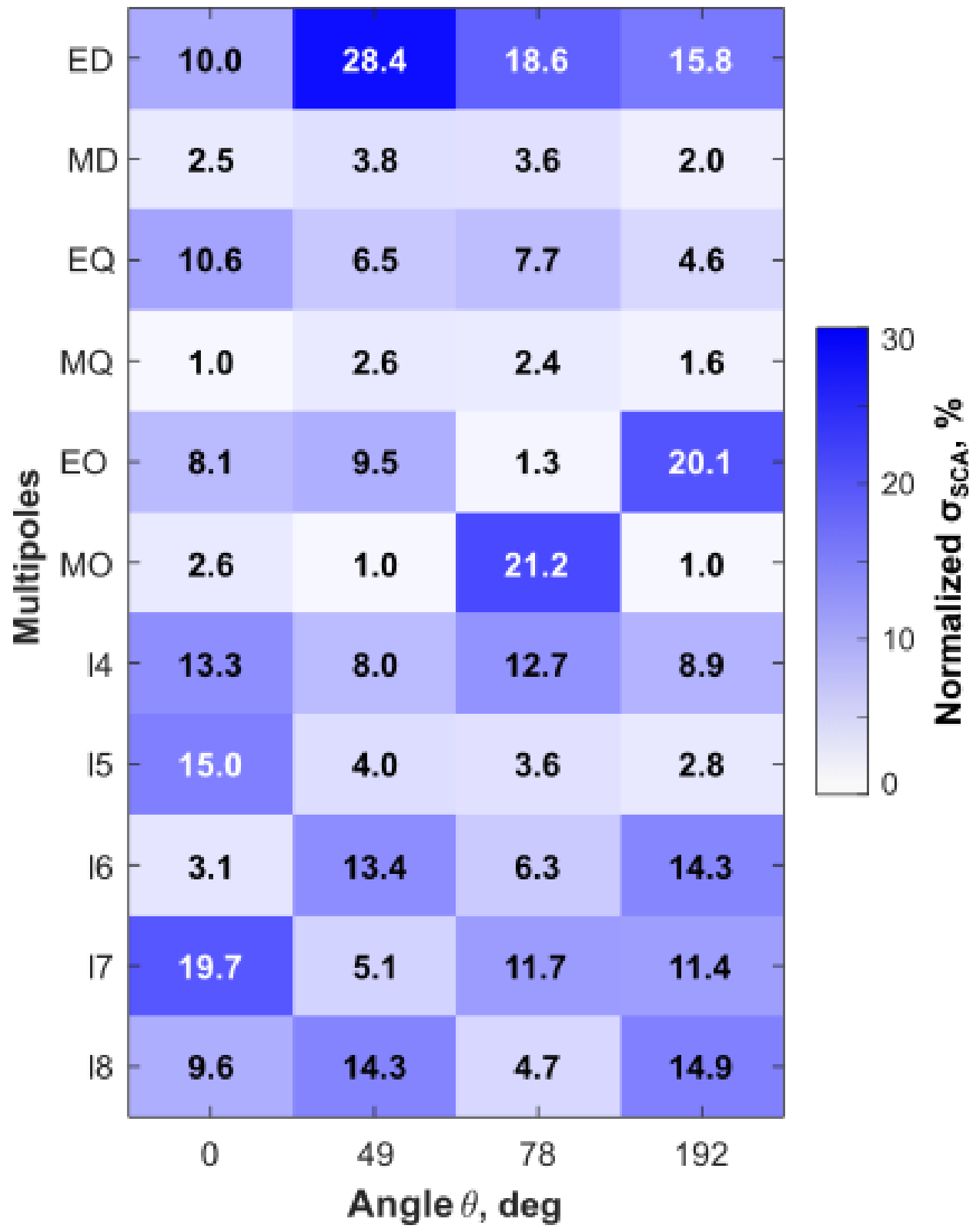

**Figure 6. Multipolar decomposition of the scattering response of the optimized 1D array.** Relative contributions of the electric dipole (ED), magnetic dipole (MD), electric quadrupole (EQ), magnetic quadrupole (MQ), electric octupole (EO), magnetic octupole (MO) and higher orders up to the 8th terms to the total scattering cross section at four representative rotation angles. The values are normalized to the total scattering at each angle.

*2D Array*

The second structure considered is a 5x5 flat array (Figure 7(a)). Compared to the 1D array, this configuration has a significantly larger number of degrees of freedom but occupies more space and spans several wavelengths. The analysis follows the same steps as for the 5x1 array. Special attention should be given to the maximum RCS, which reaches approximately 200 $cm^2$ in the end-fire direction, as shown in Figure 7(b). This value is already comparable in order of magnitude to the peak response of a small canonical trihedral reflector, even though the present structure is planar and optimized for angular modulation rather than retroreflection. Beyond the strong numerical-to-experimental agreement shown in Figure 7(b), the spectra in Figure 3(b) can now be revisited. In particular, the array exhibits a roughly fourfold increase in harmonic amplitude compared with the dipole, while the highest significant component, whose magnitude remains comparable to the dipolar response, reaches the 55th order.

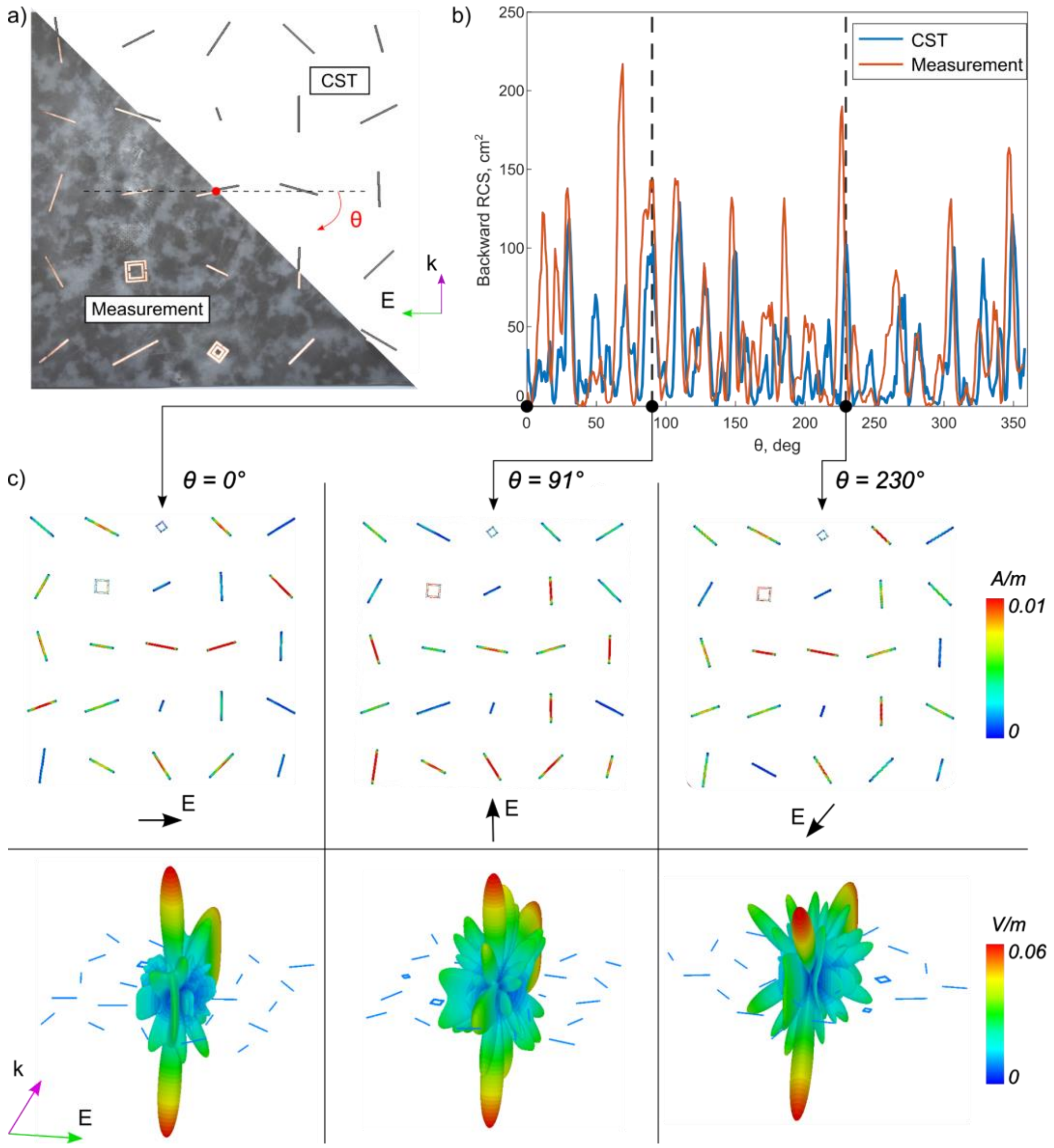

**Figure 7. Numerical and experimental characterization of the optimized 2D array. (a)** Layout of the optimized 2D array overlaid with a photograph of the fabricated scatterer. **(b)** Comparison of the angular dependence of the RCS obtained numerically and experimentally at 10 GHz. **(c)** (Top row) surface current distributions, and (bottom row) far field electric patterns, calculated at 10 GHz for representative rotation angles of 0°, 91°, and 230°.

The resonance-cascading effect is more pronounced in this case and is clearly evident in the surface current distributions shown in Figures 3(c) and 6(c). The red regions appear at different positions across the array, indicating which resonators are most strongly excited at each angle.

The scattering patterns also exhibit stronger directivity due to the larger area and the greater number of elements participating in the interaction.

A similar near-field analysis was carried out for the optimized 2D structure. The numerical and experimental maps at θ=0°, presented in Figure 8, confirm strong field localization and are consistent with the current redistribution mechanism responsible for the enhanced scattering response.

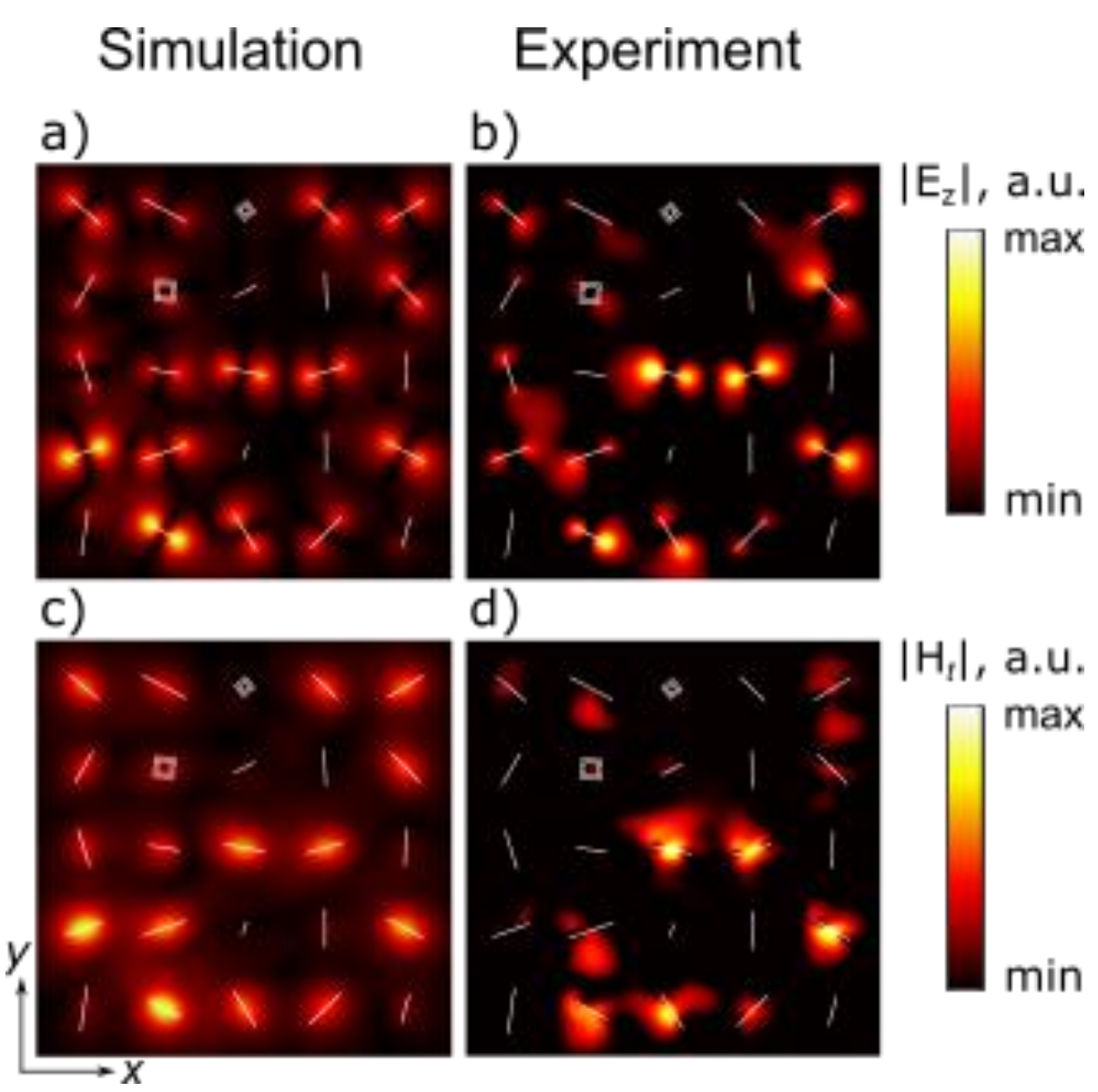

**Figure 8. Near-field electric and magnetic distributions of the optimized 2D array at θ=0°. (a,b)** Magnitude of the electric field component $|E_z|$: numerical result, left, and experimental map, right. **(c,d)** Magnitude of the magnetic field $|H_t|$: numerical result, left, and experimental map, right.

*Direct demonstration of meta-micro-Doppler*

Having established the angular scattering response and its multipolar origin, the resulting micro-Doppler signatures under rotation are now examined directly. Three samples, namely a single half-wavelength metallic dipole, the optimized 1D array, and the optimized 2D array, were rotated about their axis at an angular frequency ω=2π·4 [rad/s] in front of the antenna, as shown in Figure 9(a). Complex-valued $S_{11}$ parameters were recorded over 5 seconds using 30001 temporal points. The resulting data were then postprocessed to obtain the spectrograms shown in Figure 9(b) by applying a single-sided FFT after suppression of the zero Doppler contribution with a moving target indicator (MTI) filter, using a 0.33 s time window and 50% overlap between adjacent segments. The spectrograms show only minor temporal variations in the Doppler frequencies, indicating good stability of both the rotation and the data-acquisition processes. In addition, Figure 8(c) shows the corresponding baseband micro-Doppler combs obtained by applying the FFT to the full 5-second time trace.

The results reveal pronounced harmonic multiplication for the optimized structures. At a rotation frequency of 4 Hz, the resonant metallic dipole is mainly characterized by the second harmonic near 8 Hz, consistent with the dipolar limit. In contrast, the 1D array exhibits components up to about 30 Hz, that is, approximately the 8th harmonic, whereas the 2D array extends to about 140 Hz, or approximately the 35th harmonic. This corresponds to an increase in the highest observable harmonic order by about ×4 for the 1D array and about ×17 for the 2D array relative to the dipole. Given this level of frequency multiplication, the observed response can be considered a giant, as it shifts the rotor micro-Doppler signature well above the spectral region typically occupied by urban clutter, as discussed in the Conclusion.

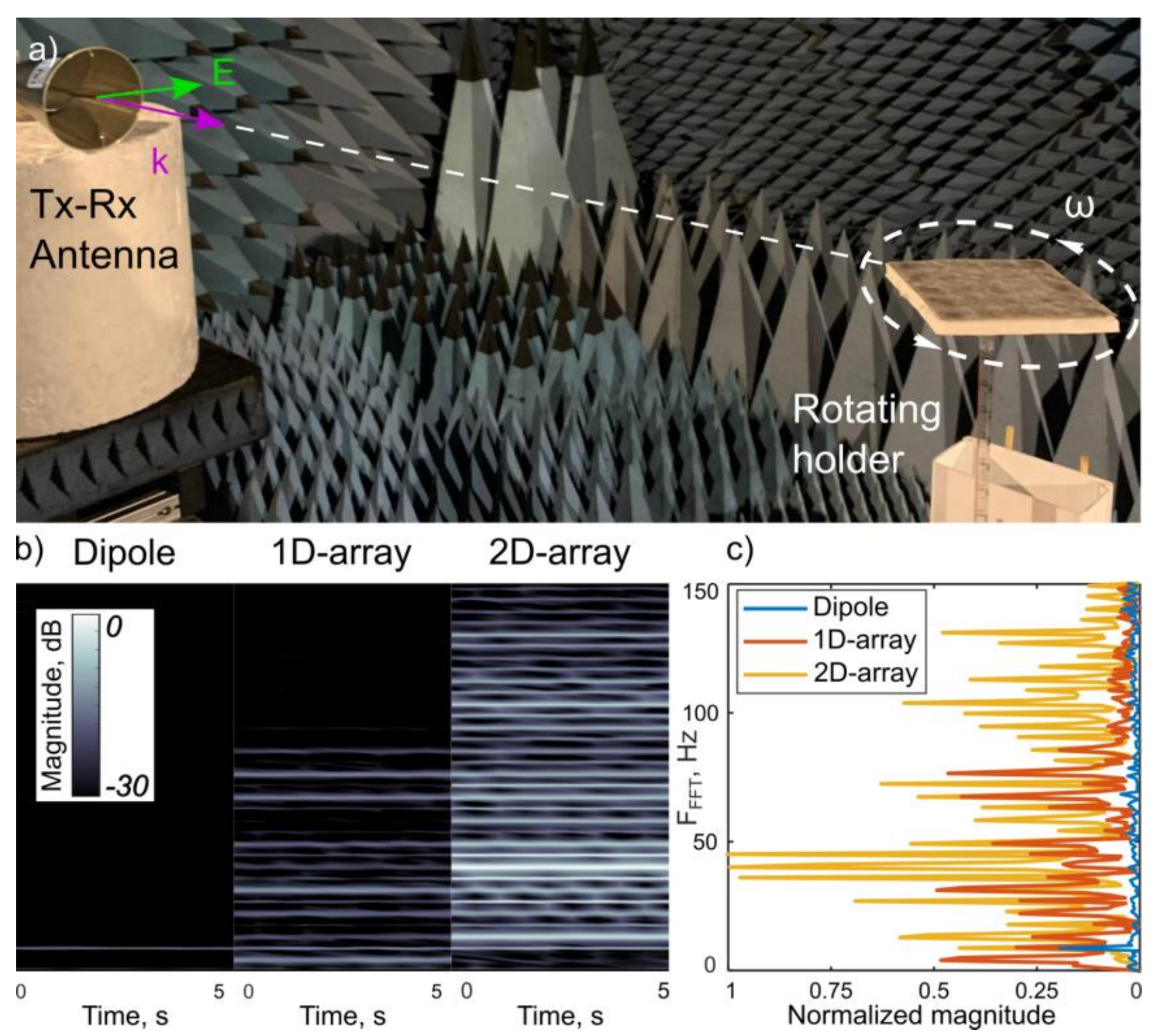

**Figure 9. Experimental observation of meta-micro-Doppler generation under rotation.** **(a)** Experimental setup for measuring the micro-Doppler response of scatterers rotating at 4 Hz. **(b)** Spectrograms recorded over 5 sec for a resonant electric dipole, the optimized 1D array, and the optimized 2D array under illumination at 10 GHz. **(c)** Corresponding baseband spectra obtained from the full-time traces.

Finally, having established the existence of a giant meta-micro-Doppler, it is important to examine how sensitive this effect is to frequency bandwidth and elevation angle. These parameters were not included in the original optimization, and, given the structure's resonant nature, the response may be narrowband and limited in angular acceptance, potentially constraining future practical applications.

Figures 10(a) and 10(b) show the dependence of the RCS of the 1D array on frequency and rotation angle θ, obtained numerically and experimentally, respectively. The purpose of this analysis is to assess the bandwidth limitations of the phenomenon. Each horizontal cut through the colormap is analogous to the angular dependence shown in Figure 2(c), with the slice at 10 GHz corresponding exactly to that result. Remarkably, the angular variation is preserved over the 7 to 12 GHz range, while within the 9 to 11 GHz window, the response retains the 10 GHz design structure particularly well. Numerical and experimental results also remain in good agreement throughout this range.

Figures 10(c) and 10(d), corresponding to the numerical and experimental results, respectively, show the RCS dependence on elevation angle under 10 GHz illumination. As seen, the colormap preserves nearly vertical features, indicating a very weak sensitivity to elevation over the full examined range of 25°. This robustness is particularly important in radar interrogation scenarios, where the rotating plane may be tilted relative to the radar antenna.

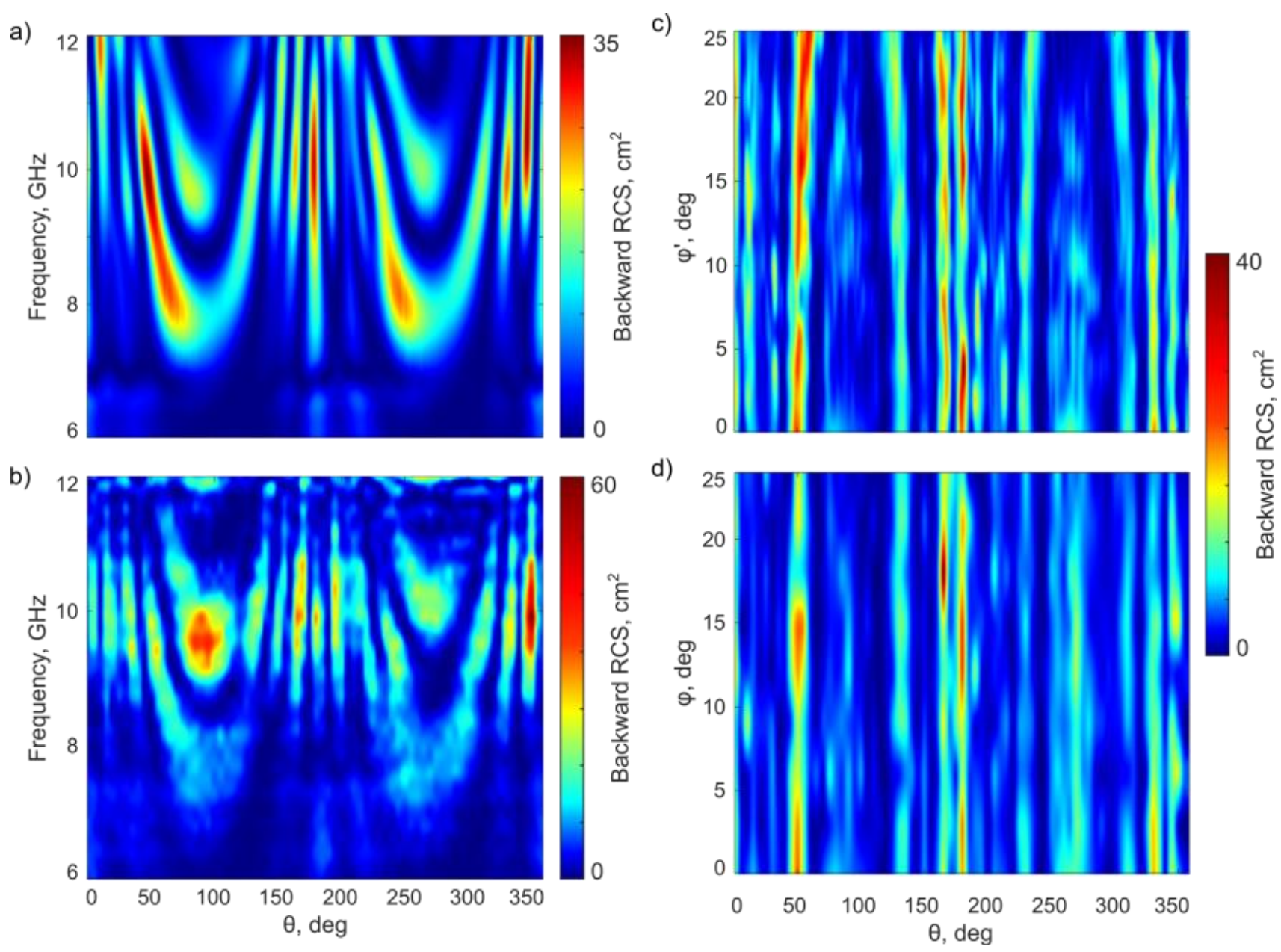

**Figure 10. Bandwidth and angular robustness of the optimized 1D array. (a,b)** Dependence of the monostatic RCS on frequency and rotation angle θ in the 6 to 12 GHz range, obtained numerically, **(a)**, and experimentally, **(b)**. Each horizontal cut is analogous to the angular response shown in Figure 2(c), with the slice at 10 GHz corresponding exactly to that result. **(c,d)** Dependence of the monostatic RCS on elevation angle under 10 GHz illumination, obtained numerically, **(c)**, and experimentally, **(d)**.

## CONCLUSION

Wireless monitoring of urban spaces is becoming increasingly important as autonomous vehicles and UAV-based traffic control systems emerge. In this context, at 10 GHz, representative urban movers such as pedestrians, slow vehicles, and many birds typically produce Doppler frequencies below ~1 kHz. A conventional drone rotor operating at about 80 to 100 Hz yields a regular micro-Doppler response near 160 to 200 Hz, which falls directly within this clutter band and therefore complicates detection. An engineered ×25 times frequency multiplication would shift the same signature to ~2.0-2.5 kHz, placing it well above the dominant urban clutter region. A similar approach may also be extended to rotating wheels and other moving mechanical parts, enhancing their radar visibility and facilitating classification.

Motivated by the above, the central problem addressed in this work is the limited frequency multiplication of rotational micro-Doppler in compact scatterers. For wavelength-scale and subwavelength objects, the response is typically governed by the dipolar channel, which confines the dominant spectral content to low harmonic orders. This challenge was addressed by introducing genetically engineered magneto-electric arrays that maximize angular-scattering contrast and promote high-order multipolar contributions to scattering. The optimization was followed by numerical and experimental characterization of 1D and 2D structures through angular RCS measurements, near-field mapping, multipolar analysis, and direct rotating tests. Together, these results revealed a resonance cascading mechanism, in which different resonant multipolar contributions dominate at different orientations and produce a strongly modulated scattering response. This engineered angular behavior was then directly converted into enhanced micro-Doppler combs during rotation. In the rotating experiments at 4 Hz, the dipolar reference was dominated by the second harmonic near 8 Hz, whereas the optimized 1D array extended to about 30 Hz, and the 2D structure reached about 140 Hz, corresponding to approximately the 8th and 35th harmonics, respectively. Given this level of frequency multiplication, the effect can justifiably be termed giant meta-micro-Doppler.

The use of metasurface elements described here is not limited to rotational tagging and radar detection. More broadly, the results establish a route toward engineered electromechanical RF transduction through time-varying resonant scattering. For example, unlike in conventional Doppler vibrometry, a metasurface integrated onto a vibrating membrane could convert acoustic motion into RF sideband signals by modulating the scattering response of an incident carrier wave. Owing to the strong dependence of engineered multipolar resonances on displacement, strain, and curvature, such structures may provide enhanced sensitivity to nanometer-scale mechanical motion compared to conventional vibrometric approaches. Beyond rotational micro-Doppler generation, these concepts may enable passive wireless acoustic sensors, smart sensing fibers and fabrics[38,39], distributed structural monitoring[40], and low-power cyber–physical interfaces[41] based on programmable electromagnetic signatures.

More broadly, the results show that rotational micro-Doppler is not merely a passive consequence of geometry, but a response that can be deliberately engineered, opening a route toward practical passive tagging, enhanced detectability, and reliable classification of rotating objects in realistic radar environments.

**Appendix 1.**

Parameters of structures

1D-array

| | **L, mm** | **α, degree** |
|---|---|---|
| **1** | 12.8 | 12.5 |
| **2** | 12.5 | -83.5 |
| **3** | 12.5 | -22.0 |
| **4** | 12.3 | 8.5 |
| **5** | 13.5 | -69.5 |

2D-array

| | **L, mm** | **α, degree** | | **L, mm** | **α, degree** |
|---|---|---|---|---|---|
| **1** | 14.3 | 81.7 | **14** | 12.2 | 16.6 |
| **2** | 11.0 | -28.5 | **15** | 11.7 | 88.1 |
| **3** | 13.0 | -57.9 | **16** | 11.2 | 60.0 |
| **4** | 13.6 | 45.0 | **17** | 5.7 | -4.1 |
| **5** | 9.2 | 75.1 | **18** | 7.4 | 27.0 |
| **6** | 12.3 | 19.8 | **19** | 11.6 | -87.3 |
| **7** | 14.4 | 19.2 | **20** | 13.0 | -45.2 |
| **8** | 4.4 | 71.0 | **21** | 12.1 | -40.4 |
| **9** | 11.9 | -89.8 | **22** | 15.2 | 27.7 |
| **10** | 12.8 | -28.0 | **23** | 3.4 | -28.0 |
| **11** | 12.5 | -72.4 | **24** | 10.6 | -52.5 |
| **12** | 9.4 | 9.8 | **25** | 12.8 | -44.1 |
| **13** | 12.9 | 11.0 | | | |